Electric-Field Modulation of Thermopower for the KTaO$_3$ Field Effect Transistors

Akira Yoshikawa[1], Kosuke Uchida[1], Kunihito Koumoto[1], Takeharu Kato[2], Yuichi Ikuhara[2,3], and Hiromichi Ohta[1,4]*

[1]*Graduate School of Engineering, Nagoya University, Furo, Chikusa, Nagoya 464-8603, Japan*

[2]*Japan Fine Ceramics Center, 2-4-1 Mutsuno, Atsuta, Nagoya 456-8587, Japan*

[3]*Institute of Engineering Innovation, The University of Tokyo, 2-11-16 Yayoi, Bunkyo, Tokyo 113-8656, Japan*

[4]*PRESTO, Japan Science and Technology Agency, Honcho, Kawaguchi, Saitama 332-0012, Japan*

*E-mail address: h-ohta@apchem.nagoya-u.ac.jp

We show herein fabrication and field-modulated thermopower for KTaO$_3$ single-crystal based field-effect transistors (FETs). The KTaO$_3$ FET exhibits field effect mobility of ~8 cm$^2$V$^{-1}$s$^{-1}$, which is ~4 times larger than that of SrTiO$_3$ FETs. The thermopower of the KTaO$_3$ FET decreased from 600 to 220 µVK$^{-1}$ by the application of gate electric field up to 1.5 MVcm$^{-1}$, ~400 µVK$^{-1}$ below that of an SrTiO$_3$ FET, clearly reflecting the smaller carrier effective mass of KTaO$_3$.

Thermoelectric energy conversion technology attracts attention to convert the waste heat into electricity.[1] Generally, the performance of thermoelectric materials is evaluated in terms of a dimensionless figure of merit, $ZT = S^2 \cdot \sigma T \cdot \kappa^{-1}$, where $Z$, $T$, $S$, $\sigma$ and $\kappa$ are a figure of merit, the absolute temperature, the thermopower, the electrical conductivity, and the thermal conductivity, respectively. Today, thermoelectric materials with $ZT > 1$, which is necessary for practical applications, are being rigorously explored, mostly using charge carrier-doped semiconductors with various doping levels because the $S^2 \cdot \sigma$ value must be enhanced according to the commonly observed trade-off relationship between two material parameters in terms of charge carrier concentration ($n$): $\sigma$ increases almost linearly with increasing $n$ until ionized impurity scattering or electron-electron scattering becomes dominant, while $|S|$ decreases with $n$. Therefore, a number of materials are usually needed to optimize thermoelectric properties.

Very recently, Sakai et al.[2] reported thermoelectric properties of several Ba-doped $KTaO_3$ single crystals ($n = 5.4 \times 10^{18} - 1.4 \times 10^{20}$ cm$^{-3}$). They found that heavy ($>10^{20}$ cm$^{-3}$) electron doping in $KTaO_3$ would provide large thermoelectric properties. $KTaO_3$ (S.G.: $Pm3m$, lattice parameter $a = 3.989$ Å) is a typical band insulator with a large band gap of ~3.8 eV.[3] Metallic conductivity in $KTaO_3$ can be obtained by the appropriate impurity doping[4] and/or the introduction of oxygen vacancies.[5-7] $KTaO_3$ exhibits very high Hall mobility of $>10^4$ cm$^2$V$^{-1}$s$^{-1}$ at 2 K.[8] Since these physical properties of $KTaO_3$ are similar to those of $SrTiO_3$,[9,10] which exhibits largest $ZT$ among transition metal oxides ($n$-type), $KTaO_3$ would be promising candidate for thermoelectric application.

In order to examine the thermoelectric properties of $S$ and $\sigma$ for $KTaO_3$, we fabricated a field effect transistor (FET) structure on a single crystal $KTaO_3$ because a FET structure on single-crystalline material would be a powerful tool in optimizing thermoelectric properties because it provides the charge carrier dependence of both $S$- and $\sigma$-values simultaneously.[11] The resultant KTaO$_3$ FET exhibits following transistor characteristics: on-off current ratio of ~10$^5$, sub-threshold swing S-factor of 1.2 Vdecade$^{-1}$, threshold gate voltage $V_{th}$ of +5.2 V, and field effect mobility $\mu_{FE}$ of ~8 cm$^2$V$^{-1}$s$^{-1}$. The thermopower $|S|$ of this KTaO$_3$ FET can be modulated from 600 to 220 µVK$^{-1}$ by the application of gate electric field up to 1.5 MVcm$^{-1}$.

Here we report the fabrication and thermopower modulation of the KTaO$_3$ FET. The schematic structure and photograph of the KTaO$_3$ FET are shown in Figs. 1(a) and 1(b), respectively. First, we treated (001) KTaO$_3$ single crystal plates (10 × 10 × 0.5 mm, SHINKOSHA) with buffered NH$_4$F-HF solution (BHF, pH = 4.5) to obtain an atomically smooth surface[12] because atomically smooth heterointerface of the gate insulator/oxide may be necessary for FET fabrication.[13] The NH$_4$F concentration was kept at 10 mol/$l$. After the BHF treatment, we obtained a relatively smooth surface with steps (~0.4 nm) and terraces [Fig. 2(b)] as compared to an untreated surface [Fig. 2(a)]. Second, 20-nm-thick metallic Ti films, which would serve as source and drain electrodes, were deposited onto the stepped KTaO$_3$ surface by electron beam (EB, no substrate heating, base pressure ~10$^{-4}$ Pa) evaporation through a stencil mask. Third, 200-nm-thick amorphous 12CaO·7Al$_2$O$_3$ (*a*-C12A7, permittivity $\varepsilon_r$ = 12) glass film was deposited by a pulsed laser deposition (PLD, ~3 Jcm$^{-2}$pulse$^{-1}$, oxygen pressure ~0.1 Pa) at room temperature (RT, no substrate heating). It should be noted that *a*-C12A7 glass would be an appropriate gate insulator for SrTiO$_3$ and KTaO$_3$ as compared to *a*-Al$_2$O$_3$.[13,14] Finally, a 20-nm Ti film was deposited by EB evaporation as described above. In order to reduce the off current, the FETs were annealed at 150 ºC in air.

Figure 2(c) shows a cross-sectional high-resolution transmission electron microscope image of the *a*-C12A7/KTaO$_3$ interface region (HRTEM, TOPCON EM-002B, acceleration voltage 200 kV, TOPCON). The featureless *a*-C12A7 (upper part) is observed, although the KTaO$_3$ layer exhibits a lattice (lower part). A broad halo is seen in the selected area electron diffraction patterns of *a*-C12A7, indicating that amorphous *a*-C12A7 film was deposited on the KTaO$_3$ layer.

Transistor characteristics of the resultant KTaO$_3$ FETs were measured by using a semiconductor device analyzer (B1500A, Agilent Technologies) at RT. The channel width (*W*) and the channel length (*L*) of the FET were 400 and 200 μm, respectively. Figure 3 summarizes typical transistor characteristics, such as transfer characteristics, field-effect mobility, sheet charge concentration, and the output of these FETs. The drain current ($I_d$) of the KTaO$_3$ FET increased markedly as the gate voltage ($V_g$) increased, hence the channel was *n*-type, and electron carriers were accumulated by positive $V_g$ [Fig. 3(a)]. Relatively large hysteresis (~1 V) in $I_d$ was also observed, most likely due to traps (~10$^{12}$ cm$^{-2}$) at the *a*-C12A7/KTaO$_3$ interface. The on-off current ratio and the S-factor were >10$^5$ and ~1.2 V·decade$^{-1}$, respectively. The threshold gate voltage ($V_{th}$), obtained from a linear fit of the $I_d^{0.5}$-$V_g$ plot [Fig. 3(b)], was +5.2 V.

Using the above measured values, we calculated the sheet charge concentration ($n_{xx}$) and the field-effect mobility ($\mu_{FE}$) of the KTaO$_3$ FETs. The $n_{xx}$ values were obtained from $n_{xx} = C_i(V_g - V_{th})$, where $C_i$ was the capacitance per unit area (51 nFcm$^{-2}$). The $\mu_{FE}$ values were obtained from $\mu_{FE} = g_m[(W/L)C_i \cdot V_d]^{-1}$, where $g_m$ was the transconductance $\partial I_d / \partial V_g$. As shown in Fig. 3(c), $\mu_{FE}$ of the FET increased drastically with $V_g$ and reached ~8 cm$^2$V$^{-1}$s$^{-1}$, which is ~25% of the RT Hall mobility of electron-doped KTaO$_3$ ($\mu_{Hall}$ ~30 cm$^2$V$^{-1}$s$^{-1}$). We also note that $\mu_{FE}$ of the KTaO$_3$ FETs were a factor of 4 greater than those of SrTiO$_3$ FETs,[11] most likely due to the difference in effective mass of the charge carrier $m_e^*$ (KTaO$_3$: 0.13 $m_0$, SrTiO$_3$: 1.16 $m_0$). Furthermore, we observed a clear pinch-off and current saturation in $I_d$ [Fig. 3(d)], indicating that the operation of this FET conformed to standard FET theory.

Then, we measured field-modulated thermopower ($S_{FE}$) of the KTaO$_3$ FET. First, a temperature difference ($\Delta T$ = 0.2–1.5 K) was introduced between the source and drain electrodes by using two Peltier devices. Then, thermo-electromotive force ($V_{TEMF}$), which is the open circuit voltage between the source and the drain electrodes, was measured during the $V_g$-sweeping. The values of $S$ were obtained from the slope of $V_{TEMF}$–$\Delta T$ plots (data not shown). Figure 4 shows $S_{FE}$–$V_g$ plots for the KTaO$_3$ FETs. The $S_{FE}$-values are negative, confirming that the channel is $n$-type. $|S|_{FE}$ gradually decreases from 600 to 220 μVK$^{-1}$ by the application of gate electric field up to 1.5 MVcm$^{-1}$, due to the fact that $n_{xx}$ increases with the $V_g$ increases. These $|S|_{FE}$ values are approximately 400 μVK$^{-1}$ lower than those for a SrTiO$_3$ FET as shown in the inset.[11] Since the value of $|S|_{FE}$ strongly depends on $m_e^*$,[16] this result reflects the fact that $m_e^*$ of KTaO$_3$ (0.13 $m_0$) is lower than that of SrTiO$_3$ (1.16 $m_0$).

In summary, we have fabricated single crystal KTaO$_3$-based field-effect transistors using amorphous 12CaO·7Al$_2$O$_3$ glass gate insulator. The resultant FET exhibit following characteristics: on-off current ratio of ~10$^5$, sub-threshold swing S-factor of 1.2 Vdecade$^{-1}$, threshold gate voltage $V_{th}$ of +5.2 V, and field effect mobility $\mu_{FE}$ of ~8 cm$^2$V$^{-1}$s$^{-1}$ (a factor of 4 greater than for SrTiO$_3$ FETs). The observed values of thermopower for the KTaO$_3$ FETs were ~400 μVK$^{-1}$ below those of SrTiO$_3$ FETs, clearly demonstrating the difference of carrier effective mass $m_e^*$ (KTaO$_3$: 0.13 $m_0$, SrTiO$_3$: 1.16 $m_0$).

**Acknowledgement** A part of this work was financially supported by Ministry of Education, Culture, Sports, Science and Technology (Nano Materials Science for Atomic-scale Modification, 20047007).

**Figure 1** (a) The schematic device structure and (b) a photograph of the $KTaO_3$ FET. Ti films (20-nm-thick) are used as the source, drain and gate electrodes. A 200-nm-thick $a$-C12A7 film is used as the gate insulator. Channel length ($L$) and channel width ($W$) are 200 and 400 μm, respectively.

**Figure 2** Topographic AFM images of (a) untreated and (b) BHF-treated $KTaO_3$ single-crystal surface. A smooth surface with steps and terraces is observed in (b). (c) Cross-sectional HRTEM image of the 200-nm-thick $a$-C12A7/$KTaO_3$ heterointerface.

**Figure 3** Typical transistor characteristics of a $KTaO_3$ FET with 200-nm-thick $a$-C12A7 ($\varepsilon_r$ =12) gate insulator at RT [(a) Transfer characteristic ($I_d$ -$V_g$ plot), (b) $I_d^{0.5}$-$V_g$ plot, (c) field-effect mobility ($\mu_{FE}$)-sheet charge density ($n_{xx}$) - $V_g$ plots, (d) Output characteristic ($I_d$ –$V_d$ plot)].

**Figure 4** Field-modulated thermopower ($S$) for the $KTaO_3$ FET channel. $S$ for the $SrTiO_3$ FET channel[11] is also plotted in the inset for comparison. Thermopower |$S$| of the $KTaO_3$ FET is roughly ~400 μVK$^{-1}$ smaller than that of $SrTiO_3$ FET and can be tuned from 600 to 220 μVK$^{-1}$.

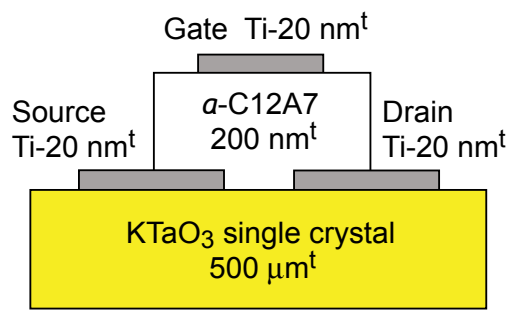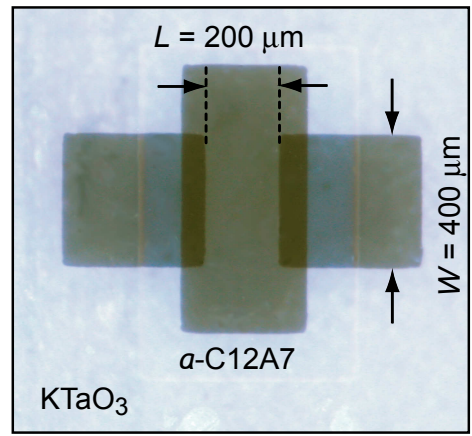

Figure 1

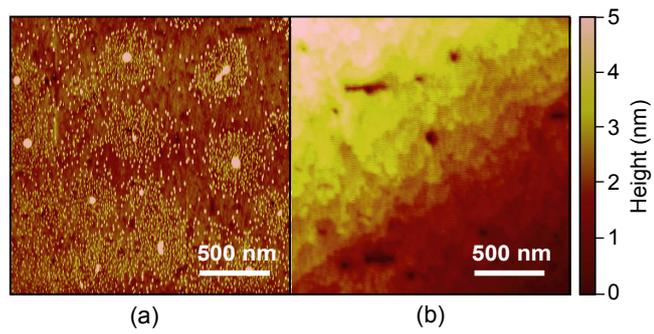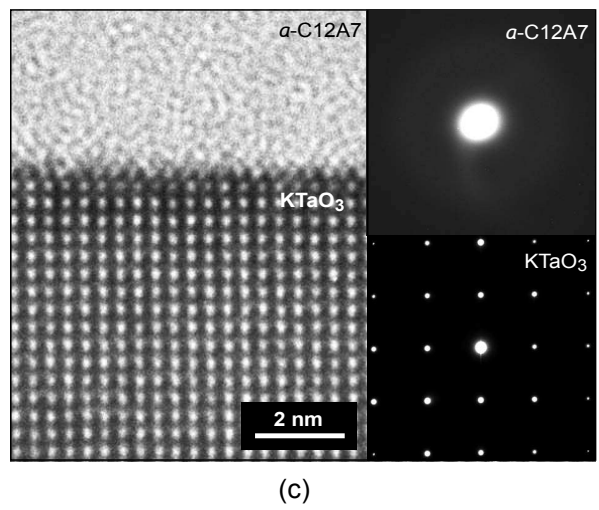

Figure 2

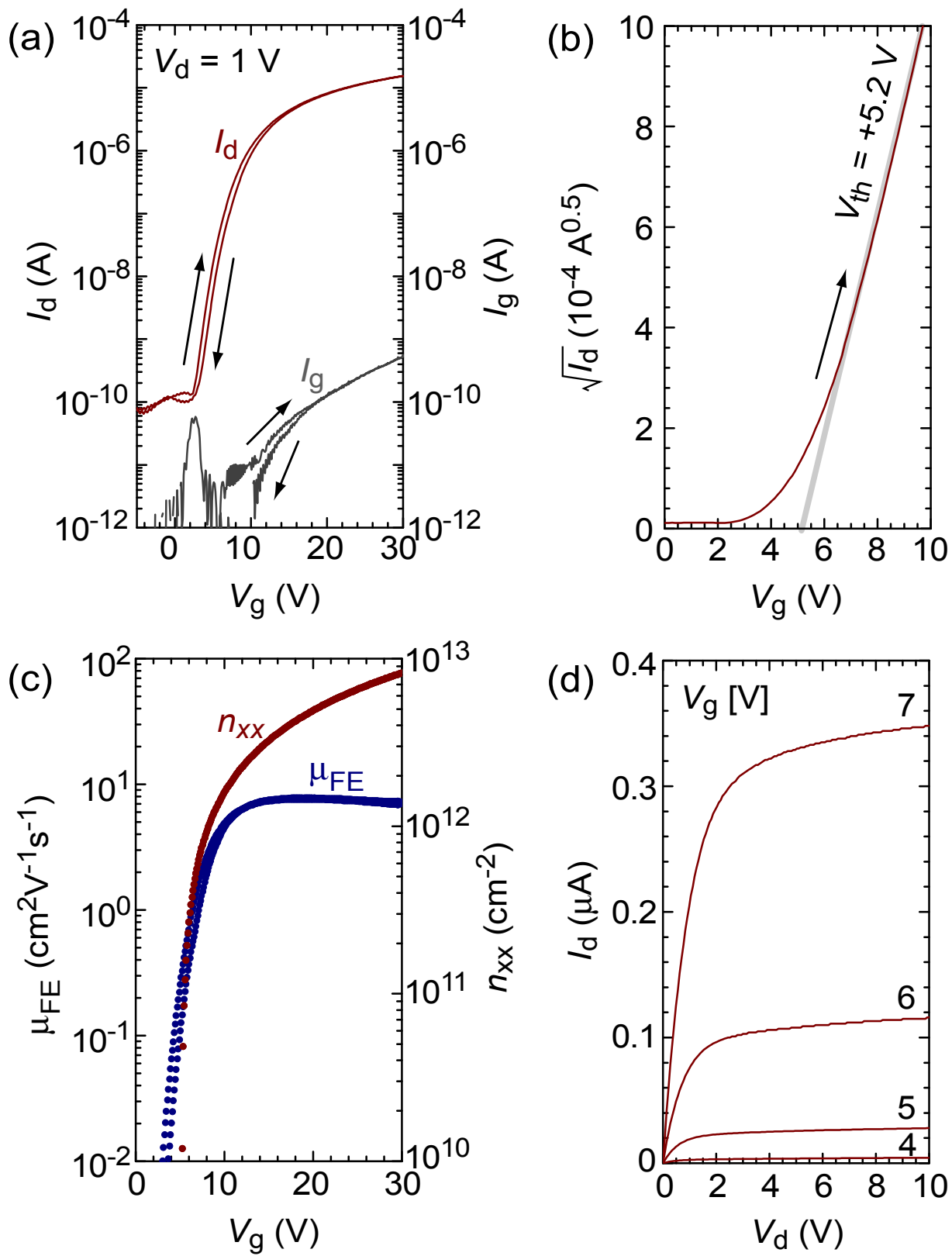

Figure 3

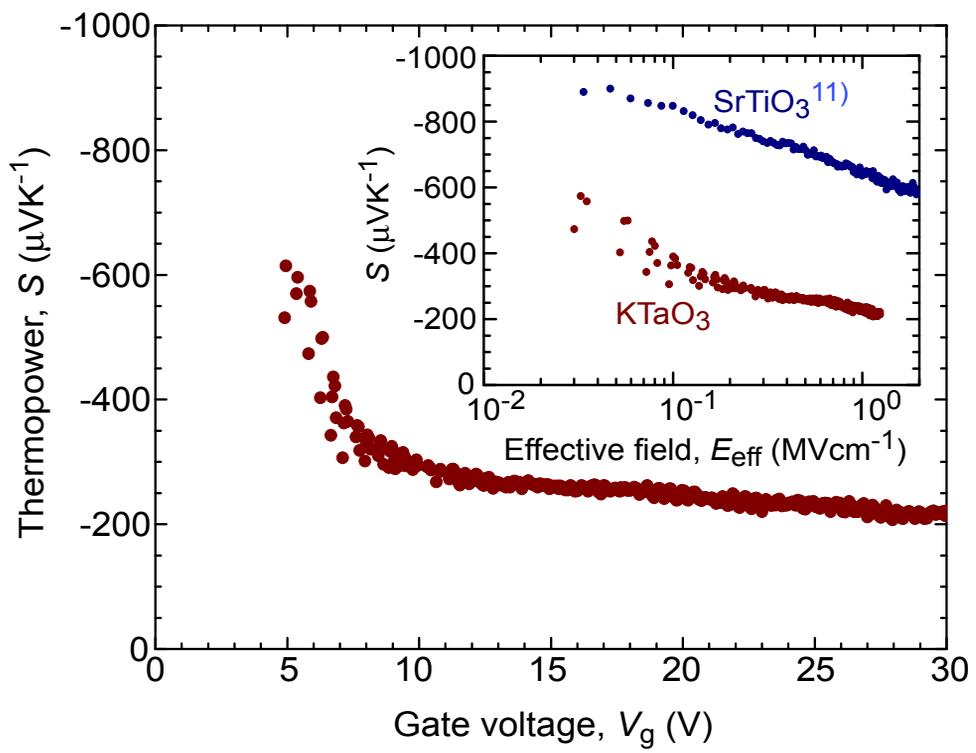

Figure 4